\documentclass[a4paper,11pt]{article}
\usepackage{jcappub} 
\usepackage{lineno}
\usepackage[makeroom]{cancel}

\usepackage{booktabs,threeparttable,makecell,array}
\usepackage[table]{xcolor}
\usepackage{pifont}
\usepackage{arydshln} 
\usepackage[normalem]{ulem}
\usepackage{tabularx}

\definecolor{Hbar}{rgb}{0.95,0.98,0.94}

\newcolumntype{L}{>{\raggedright\arraybackslash}p{3.8cm}}
\newcolumntype{C}{>{\centering\arraybackslash}c}
\newcolumntype{F}{>{\centering\arraybackslash}p{1.2cm}}
\newcolumntype{G}{>{\columncolor{Hbar}\centering\arraybackslash}c}
\newcolumntype{R}{>{\centering\arraybackslash}c}

\definecolor{MixRow}{rgb}{0.96,0.98,0.95}

\usepackage{hyperref}
\usepackage{cleveref}
\usepackage{physics}
\usepackage{tikz}
\usepackage{multirow}
\usepackage{rotating}
\usepackage{upgreek}
\usepackage{graphicx}
\usepackage{etoolbox}
\usepackage{siunitx}
\usepackage{xcolor}
\usepackage{subcaption}
\usepackage{enumitem}

\usepackage{extarrows}
\usepackage{amsmath}
\usepackage{dcolumn}
\usepackage{simpler-wick}
\usepackage{kantlipsum}
\usepackage{float}
\usepackage{rotating}

\usepackage{etoolbox}
\usepackage{extarrows} 
\usepackage{bm}
\usepackage{wasysym}
\usepackage{verbatim}

\usepackage{verbatim}
\usepackage{color}
\usepackage{mathtools,slashed}

\newcommand{\PT}{\text{PT}}
\newcommand{\mx}{\text{max}}

\crefname{equation}{}{}

\def\d{\mathrm{d}}

\def\C{\mathcal{C}}

\def\vec{\mathbf}

\def\e{\mathrm{e}}

\def\veck{\vec{k}}
\def\vecp{\vec{p}}

\arxivnumber{2601.xxxxx} 
\title{\boldmath Phenomenology of Vector Dark Matter produced by a First Order Phase Transition}

\author{M. Fairbairn,}
\author{W. S. A. Shellard}
\affiliation{Theoretical Particle Physics and Cosmology, King’s College London, Strand, London WC2R 2LS, United Kingdom}

\emailAdd{william.shellard@kcl.ac.uk}

\abstract{Both scalar and vector dark matter can be produced during a cosmological first order phase transition if the dark matter is coupled to the field undergoing the transition. Both kinds of particle are also produced by the plasma through the normal freeze out scenario. For different dark matter masses, we identify the regions of parameter space where there are significant deviations from the normal freeze out scenario and discover there are some rather general predictions. For dark matter particles in the traditional thermal relic GeV-TeV window, dark sector phase transitions around a GeV affect scalar dark matter and dark sector phase transitions around 10 MeV affect vector dark matter abundances (and therefore should take place in a dark sector). When the phase transitions are in the interesting temperature range, the normal range of dark matter masses are different to those predicted by thermal freeze out. We calculate the expected gravitational wave signal of these phase transitions.}

\begin{document}
\maketitle
\flushbottom

\section{Introduction}
\label{sec:intro}

Multiple lines of evidence suggest the presence of a large component of the Universe with a matter equation of state ($w=0$). This component is constrained to be very weakly-interacting with visible baryonic matter, hence its name, `dark matter'. This unseen component of the Universe fits into the standard $\Lambda$CDM model of cosmology.

As of yet, the exact nature of dark matter is unknown. A great many models have been put forward, which satisfy the general astrophysical constraints, but vary greatly in their physical description of dark matter -- \cite{Cirelli:2024} provides an extensive review. Most models posit some new particle that weakly interacts with the Standard Model; a particularly simple example involves the dark matter at early times being in equilibrium with the Standard Model plasma, before freezing out as a thermal relic -- the so called WIMP (weakly-interacting massive particle). However, many WIMP models are now heavily constrained by direct and indirect detection experiments, and CMB observations \cite{Arcadi_2018}. This motivates the consideration of less minimal models of dark matter, such as the one presented in this paper.

First-order phase transitions occur when a field acquires a vacuum expectation value (vev) after fluctuating across a barrier in its potential, creating a phase of matter where the original symmetry of the theory is broken. Once `bubbles' of broken phase are nucleated, they propagate outward until the entire Universe is in the broken phase. The Standard Model Higgs symmetry breaking is {\it not} an example of a first-order phase transition as the vev smoothly varies from zero to the low-temperature expectation value as the Universe cools. However, beyond the standard model modifications to this potential can introduce a suitable potential barrier.

First-order phase transitions are a topic of current interest in physics, due to their ability to provide the necessary conditions for baryogenesis \cite{Kuzmin:1985mm,Cohen:1993nk}, primordial magnetic fields \cite{Baym:1995fk,PhysRevD.37.2743}, primordial black holes \cite{Garriga:2015fdk,Liu:2021svg,Gross:2021qgx,Baker:2021nyl,Kawana:2021tde,Jung:2021mku,Lewicki:2024ghw,Ai:2024cka},\footnote{Recntly, it has been pointed out that PBH formation from curvature perturbations generated from a slow FOPT may be rather inefficient \cite{Franciolini:2025ztf,Wang:2026zvz}.} and stochastic gravitational wave signals of that could be detected by next-generation gravitational wave experiments \cite{Hogan:1986qda,Kamionkowski:1993fg,Grojean:2006bp}.

The model of dark matter we now consider emerged from a consideration of the dynamics of bubble walls in first-order cosmological phase transitions. Several mechanisms for generating dark matter in this context have been explored, notably from bubble collisions \cite{Watkins:1991zt,Konstandin:2011ds}, filtering effects \cite{Baker:2019ndr,Chway:2019kft} and bubble expansion \cite{Azatov:2021ifm,Azatov:2022tii,Baldes:2022oev,Ai:2024ikj}, which is the focus of this report. In brief, the bubble wall locally creates an environment where momentum in the direction of its propagation is not conserved, allowing for particle interactions that would usually be energetically forbidden. This allows us to create dark matter of mass much greater than the temperature at which the phase transition occurs.  For such a first order phase transition to be strong, we need to avoid too much friction.  In this work we therefore restrict ourselves to the situation where the dark sector phase transition is coupled only very weakly to the standard model degrees of freedom, but where the dark sector has the same temperature as our visible sector, suggesting that the two sectors were in equilirium at some higher temperature.

In previous work, scalar \cite{Azatov:2021ifm} and vector \cite{Azatov:2024crd,Ai:2024ikj} dark matter candidates have been considered. In this paper we examine the phenomenology of this DM production scenario in more detail, searching for characteristic signatures which may provide evidence for it, or constrain it. The structure of this paper is as follows, in Section \ref{sec:dmprod} we will discuss the production of dark matter from expanding bubble walls, including the wall's equation of motion. We then go on to consider the subsequent evolution of the dark matter in Section \ref{sec:evolution}, and  finally we turn to the possibility of the production of detectable gravitational waves \ref{sec:gw} before making our conclusions. Details of the calculations used in Section \ref{sec:dmprod} are given in Appendix \ref{sec:wallfric} and the possibility of observational effects such as inhomogeneities in the production rate is discussed in Appendix \ref{sec:obs}.

\section{Producing Dark Matter from Expanding Bubble Walls}
\label{sec:dmprod}
\subsection{Wall Equation of Motion}
\label{sec:wallEoM}
In this paper we compare two alternative models for dark matter, a massive vector $V$ and a scalar $\chi$, coupled to a scalar field which undergoes a first-order phase transition $\Phi$, giving the respective Lagrangians
\begin{align}
    \mathcal{L}_V &= \frac{1}{2}\partial_\mu\Phi\partial^\mu\Phi - V(\Phi) - \frac{1}{4}F_{\mu\nu}F^{\mu\nu} + \frac{1}{2}m_V^2 V_\mu V^\mu + \frac{\lambda}{4}\Phi^2V_\mu V^\mu\,,\\
    \mathcal{L}_\chi &= \frac{1}{2}\partial_\mu\Phi\partial^\mu\Phi - V(\Phi) + \frac{1}{2}\partial_\mu\chi\partial^\mu\chi - \frac{1}{2}m_\chi^2 \chi^2 + \frac{\lambda}{4}\Phi^2\chi^2\,.
\end{align}
We remain ambivalent about the exact form of $V(\Phi)$ and simply characterise it by the vev of the broken phase $v_b$; the zero-temperature potential difference between the broken and symmetric phases $\Delta V$; the temperature of the phase transition $T_\text{PT}$; and the inverse transition duration $\beta$. We then substitute $\Phi(x) \rightarrow v(x) + \phi(x)$, which generates an additional mass term for the dark matter, $\sqrt{\lambda v_{\smash b}^2 / 2}$, which we may ignore (we will see that the existing mass term is much larger than that acquired from the vev), and the interaction term
\begin{equation}
    \mathcal{L}_V \supset \frac{\lambda}{2}v(x)\phi V_\mu V^\mu\,,\quad\mathcal{L}_\chi \supset \frac{\lambda}{2}v(x)\phi \chi^2\,.
\end{equation}
in the vector and scalar cases respectively. When the field undergoes a first-order phase transition, bubbles of the new phase nucleate and propagate outwards until the whole universe is in the new phase. We take the following Ansatz for the profile of a bubble wall:
\begin{equation}
    v(r) = \frac{v_b}{2}\left[1 + \tanh{\frac{r - r_w(t)}{L_w/\gamma_w}}\right]\,,
\end{equation}
where $L_w$ is the width of the wall in its own instantaneous rest frame, and $\gamma_w$ is its Lorentz factor. Following \cite{Gouttenoire:2023naa}, the equation of motion for the bubble wall reads
\begin{equation}
    \dot{\gamma}_w + 3H v_w^2\gamma_w + \frac{2}{r_w}v_w\gamma_w =\frac{3L_w}{v_b^2}(\mathcal{P}_\text{driving} - \mathcal{P}_\text{friction})v_w\,,
\end{equation}
with the velocity of the wall taken to be $v_w\approx 1$ henceforth. The driving force $\mathcal{P}_\text{driving}$ is the zero-temperature potential difference between the symmetric and broken phases $\Delta V$, so is a constant. The friction $\mathcal{P}_\text{friction}$ is caused by particles interacting and changing in mass as they cross the wall, the exact form of which will be calculated in the following section. Sufficient for now is that the friction in the vector case goes as $\mathcal{P}_\text{friction} = \mathcal{C}_V\cdot \gamma_w^2$ and in the scalar case it is $\mathcal{P}_\text{friction} = \C_\chi \cdot\ln\gamma_w$ at large values of $\gamma_w$. As at percolation, the bubble radius will only be some fraction of the Hubble radius, the Hubble friction term will always be smaller than the curvature term, so we neglect it. In the vector case, exchanging $r_w$ for $t$, as $v_w\approx 1$,
\begin{equation}
    \gamma_w(r_w) = \sqrt{\frac{\Delta V}{\mathcal{C}}}\coth\left(\frac{3 L_w \sqrt{\mathcal{C}\Delta V}}{v_b^2}\cdot r_w\right) - \frac{v_b^2}{3 L_w \mathcal{C} \cdot r_w}\,.
\end{equation}
The bubble thus rapidly accelerates before reaching a terminal velocity $\gamma^\text{max}_w$. The scalar case does not have an analytical solution but exhibits a similar behaviour. A lengthscale for the distance the bubble propagates before reaching its terminal velocity $r_\text{term}$ is found by dividing $\gamma^\text{max}_w$ by the initial gradient of the solution, giving
\begin{equation}
    r_\text{term} = \frac{\gamma_w^\text{max} v_b^2}{3\Delta V L_w}\,.
\end{equation}

\begin{figure}[b]
    \centering
    \includegraphics[scale=0.8]{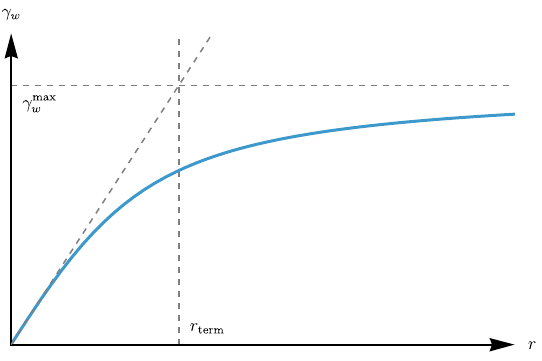}
    \caption{Solution for the wall velocity in the vector case with means of calculating the speed-up radius of the wall $r_\text{term}$ depicted.}
    \label{fig:wallsoln}
\end{figure}

\subsection{Wall Friction and Produced Dark Matter}
\label{sec:friction}
As particles cross the bubble wall, they may undergo various processes that extract momentum from the bubble wall. We may simplify matters by stating that in the ultrarelativistic regime, the wall is moving sufficiently fast that two-particle scattering events over the wall become negligible relative to $1\rightarrow n$ processes; this is known as the ballistic approximation \cite{Liu:1992tn,Bodeker:2017cim,BarrosoMancha:2020fay,Ai:2024btx}. Furthermore, we also assume we are in the detonation regime, where particle reflections from the bubble wall are negligible and thus the plasma in front of the wall can be treated as unperturbed.

Previous calculations have demonstrated the existence of a friction term arising from particles changing mass \cite{Bodeker:2009qy}, or emitting a gauge boson as transition radiation \cite{Bodeker:2017cim} as they cross the wall. The former approaches a constant value for large wall velocities, while the latter grows linearly with $\gamma_w$, and thus strongly restricts how fast the bubble wall may travel. In order to achieve efficient production of dark matter, we require it to be the dominant source of friction on the bubble wall, so that the majority of the vacuum energy released during the phase transition goes into the dark matter production mechanism. For this reason, we consider a transition in a dark sector, where the field undergoing the phase transition $\phi$ only couples to dark matter (and possibly also the Higgs), allowing us to neglect transition radiation and friction from changes in mass except from that of $\phi$ itself (we assume that the dark matter density is thermally suppressed before the phase transition), which can be neglected if we impose
\begin{equation}
    \Delta m_\phi \ll \frac{\sqrt{24\,\Delta V}}{T_\text{PT}}\,.
\end{equation}
In free space, the process $\phi\rightarrow VV$ or $\phi\rightarrow \chi\chi$ is energetically forbidden but, working in the frame of the wall, the wall breaks $z$-translation invariance, allowing for incoming $\phi$ particles to exchange some $z$-momentum $\Delta p^z$ for rest-mass energy and produce two dark matter particles. The wall then absorbs the difference in momentum and experiences a backwards pressure. In Appendix \ref{sec:wallfric} this calculation is done explicitly in the large $\gamma_w$ regime, in which the integral becomes analytically tractable, giving the friction in the vector case
\begin{equation}
    \mathcal{P}_{\phi\rightarrow  VV} \approx \frac{3\,\zeta(3)}{8 \pi^6} \frac{T_\text{PT}^4 v_b^2 \lambda^2}{L_w^2 m_V^4}\gamma_w^2\,,
\end{equation}
where the numerical prefactor $\approx 4.7\times10^{-4}$. This confirms the quadratic scaling in $\gamma_w$ found by numerical integration in \cite{Ai:2024ikj} and extends it to arbitrary values of $\gamma_w$, at least until unitarity violation becomes a concern for the Proca field we are considering. Note that while this approximate analytical result has the same form as the numerical result found in \cite{Ai:2024ikj}, it differs in its prefactor by a factor of about 2. This discrepancy is much more dramatic in the prefactor of $n_V$ below, which is a factor of $10^3$ larger in this paper than in the numerical result, due to an error in the previous work.\footnote{This error will be corrected in an erratum of Ref. \cite{Ai:2024ikj}.} With an expression for the friction, we find that
\begin{align}
    \gamma_{w,V}^\text{max} = \sqrt{\frac{8\pi^6}{3\zeta(3)}} \frac{L_w m_V^2 \sqrt{\Delta V}}{T_\text{PT}^2 v_b \lambda}\,.
\end{align}
The produced dark matter number density is then saturated at $\gamma_{w,V}^\text{max}$ and is given by
\begin{equation}
    n_V \approx \frac{T_\text{PT}^4 v_b^2 \lambda^2}{24 \pi^3 L_w m_V^4} \cdot \gamma_w\,,\quad
    n_V^\text{max} = \frac{T_\text{PT}^2v_b^2\sqrt{\Delta V} \lambda}{6\sqrt{6} \zeta(3) m_V^2}\,.
\end{equation}
In the scalar case we obtain
\begin{equation}
    \mathcal{P}_{\phi\rightarrow \chi\chi} = \frac{T_\text{PT}^2 v_b^2 \lambda^2}{128\pi^4}\left(-1-\gamma_E+\log\left(\frac{4T_\text{PT}\gamma_w}{\pi L_w m_\chi^2}\right)\right)\,.
\end{equation}
\begin{equation}
    n_\chi = \frac{T_\text{PT}^3 v_b^2 \lambda^2}{48\pi^4 m_\chi^2}\,.
\end{equation}
The numerical results have an additional exponential factor, but this easily saturates for generic parameter values, thus the number density is independent of the wall velocity.

It is also worth knowing the average momentum per produced dark matter particle immediately after their production in both cases. This is (Appendix \ref{sec:wallfric})
\begin{equation}
    p_V \approx \langle E_{V}\rangle = \frac{\mathcal{P}_{\phi\rightarrow  VV}}{n_V} = \frac{9\zeta(3)}{\pi^3}\gamma_w L_w\,,\quad
    p_\chi \approx \frac{3m_\chi^2}{8T_\PT}\left(-1-\gamma_E+\log\left(\frac{4T_\text{PT}\gamma_w}{\pi L_w m_\chi^2}\right)\right)\,.
\end{equation}
At the wall's terminal velocity, we have
\begin{equation}
    p_V^\text{max} = 6\sqrt{6 \zeta(3)}\frac{m_V^2\sqrt{\Delta V}}{T_\text{PT}^2 v_b \lambda}\,.
\end{equation}
Moving forward, we will re-express results as follows: the phase transition strength parameter $\alpha$ may be used to relate $\Delta V$ and $T_\text{PT}$ through its definition as the ratio of the false vacuum energy to the Standard Model plasma energy:
\begin{equation}
    \alpha = \frac{\Delta V}{\rho_\text{rad}} = \frac{\Delta V}{\frac{\pi^2}{30}g_{*}(T_\text{PT})T_\text{PT}^4}\,,
    \label{eq:alpha}
\end{equation}
and, as in the absence of any fine tuning $v_b$ and $T_{PT}$ should be related to each other by order unity dimensionless values, we will therefore take
\begin{equation}
    v_b = \xi T_\text{PT}\,.
    \label{eq:xi}
\end{equation}
In addition, for the remainder of this paper we shall take $g_{*S}(T_\text{PT}) \approx g_{*}(T_\text{PT}) \approx g_*(m_V) \approx g_{*}$, and as we are considering a fast-moving wall, the phase transition will complete quickly, and we take the nucleation and percolation temperatures to be equal, expressed as $T_\text{PT}$.

\subsection{Producing the Correct Relic Density}
\label{sec:relic-dens}

At creation, the dark matter has a density of $n_\text{PT}$. If we assume that the bubble is travelling at its terminal velocity for most of its propagation (as will be justified in Appendix \ref{sec:obs}), and that the phase transition is responsible for all of the dark matter in the universe (freeze-out production will be considered in Sec. \ref{annihilations}), and that no dark matter annihilations occur (Sec. \ref{annihilations}), the number of dark matter particles per comoving volume
\begin{equation}
    Y_\text{PT} = \frac{n_\text{PT}}{s_\text{PT}}
    = \frac{45}{2\pi^2}\frac{n_\text{PT}}{g_{*}T_\text{PT}^3}\,,
\end{equation}
is conserved, where $s_\text{PT}$ is the entropy density at the phase transition. The dark matter relic abundance $\Omega_\text{DM0}h^2 = 0.121$, so we obtain the constraint ($\zeta=100$\,km\,s$^{-1}$Mpc$^{-1}$)
\begin{equation}
    \Omega_\text{DM0}h^2 = 0.121 = \frac{\rho_\text{DM0}}{\rho_c}h^2
    = \frac{8\pi G s_0 m_V Y}{3\zeta^2}\,.
    \label{relic_density}
\end{equation}
Rearranging, for the vector case,
\begin{equation}
    T_\text{PT} = 2.30\times \bigg(\frac{g_{*}}{\alpha \xi^2}\bigg)^{\!\frac{1}{4}} \bigg(\frac{0.1}{\lambda}\bigg)^{\!\frac{1}{2}} \bigg(\frac{m_V}{100\,\text{GeV}}\bigg)^{\!\frac{1}{2}}\,\text{MeV}\,.
    \label{Tprod}
\end{equation}
For the scalar case,
\begin{equation}
    T_\text{PT} = 0.94\times \frac{1}{\xi} \bigg(\frac{g_*}{100}\bigg)^\frac{1}{2}\bigg(\frac{0.1}{\lambda}\bigg)\bigg(\frac{m_V}{100\,\text{GeV}}\bigg)^\frac{1}{2}\,\text{GeV}\,.
\end{equation}

\section{Subsequent Dark Matter Evolution}
\label{sec:evolution}

After creation, the dark matter may annihilate with itself, potentially leading to a significant decrease in its density, requiring us to refine the simple picture described above. We also have seen that the dark matter particles are created with a large momentum $p_V$, which is in the direction of the bubble's propagation. Therefore we must consider the interactions with the $\phi$-plasma, which limit how the dark matter can free-stream before thermalising, and the self-interactions of the dark matter which cause it to annihilate.

\subsection{Scatterings and Rethermalisation}

We first estimate how rapidly these particles slow down and thermalise through collisions with the $\phi$-particle plasma. Key to this is understanding how much momentum a dark matter particle loses in a given collision. Transforming to the centre-of-momentum frame of a colliding $V$/$\chi$ and thermal $\phi$, neglecting the mass $m_{V/\chi}$, the momentum becomes $p_{\text{CoM}}^z\approx p^z/2\gamma_{\text{CoM}}$. After the collision this momentum may be deflected in any direction, i.e. $p_{\text{CoM}}'^{z}=p_{\text{CoM}}^{z}\cos{\theta}$, which transforms back to the plasma frame as $p'\approx \tfrac{1}{2}(1+\cos\theta)$, so the produced dark matter may lose up to all of its momentum in a single collision. The relaxation rate to kinetic equilibrium is given by
\begin{align}
\label{eq:relax-rate}
    \Gamma_{\rm relax} &= n_\phi \sigma(\phi V\rightarrow\phi V) v_{\rm rel}\frac{\delta p_V}{p_V}
\end{align}
If we take $n_\phi = (\zeta(3)/\pi^2) T_\text{PT}^3$, i.e. the $\phi$-plasma is still relativistic, $\sigma(\phi V\rightarrow\phi V) \approx 9\lambda^2 / (64\pi p_V T_\text{PT})$, and $\delta p_V / p_V \approx 1/2$, then we find that
\begin{equation}
    t_\text{relax} H = \frac{H}{\Gamma_\text{relax}} = 4.91\times10^{-8} \times \sqrt{\alpha}\bigg(\frac{m_V}{100\,\text{GeV}}\bigg)^2\bigg(\frac{T_\text{PT}}{\text{GeV}}\bigg)\bigg(\frac{0.1}{\lambda}\bigg)^3.
    \label{trelax}
\end{equation}
So the produced dark matter will very quickly impart its large initial momentum into the plasma, relative to the Hubble time. The scalar version of this calculation is identical, except without the factor of 9 in the cross section from summing over polarisation states.

\subsection{Annihilations}\label{annihilations}

Now, to consider annihilations of dark matter, considering here the vector case, we start with the integrated Boltzmann equation
\begin{align}
    \frac{dY}{dx} &= -\frac{\eta}{x^2}[Y(x)^2 - Y_\text{eq}(x)^2]\,,\\
    \eta &= \frac{s\langle\sigma v\rangle}{H}\bigg\rvert_{T=m_V},
\end{align}
where $Y = n_V/s$ and the dimensionless variable $x = m_V/T$, and we have assumed that the dark matter has rethermalised with the plasma. This assumption also allows us to take the thermally-averaged cross-section of dark matter annihilations to be
\begin{equation}
    \langle\sigma v\rangle \big\rvert_{m_V\gg T} \approx
    \frac{9\lambda^2}{64\pi m_V^2}\,.
\end{equation}
The scalar cross section omits the factor of 9. If the phase transition occurs before the ordinary freeze-out of the dark matter from the thermal bath, then the dark matter density will quickly re-equilibrate and the phase transition will have made no impact on the final amount of dark matter in the late universe. The phase transition should therefore occur after thermal freeze-out of the dark matter i.e. $x_\text{PT} > x_\text{FO}$, where we may calculate $x_\text{FO}$ iteratively using
\begin{equation}
    x_\text{FO} = \ln\left(\frac{3\sqrt{90}}{256\pi^4}\frac{\lambda^2}{\sqrt{g_{*}\,x_\text{FO}}}\frac{M_\text{Pl}}{m_V}\right)\,.
\end{equation}
We will neglect the equilibrium abundance term $Y_\text{eq}$ as after freeze-out this is much less than the true abundance. It is now straightforward to integrate the Boltzmann equation, giving the solution
\begin{equation}
    Y(\infty) = \left(\frac{\eta}{x_\text{PT}} + \frac{1}{Y(x_\text{PT})}\right)^{-1}\,.
\end{equation}
We can then identify two extremes in the behaviour of the dark matter after it is created. Firstly, as in \cite{Ai:2024ikj}, when $\eta/x_\text{PT} \gg Y(x_\text{PT})^{-1}$, the large amount of dark matter created reignites annihilations, reducing the overall abundance to $Y(\infty) = x_\text{PT}/\eta$, where we now notice that the dark matter density immediately after production doesn't impact the final abundance. Alternatively, the coupling is weak enough for annihilations to remain disfavoured after the phase transition i.e. $\eta/x_\text{PT} \ll Y(x_\text{PT})^{-1}$, as in Sec. \ref{sec:relic-dens}, and the comoving dark matter density remains constant: $Y(\infty) = Y(x_\text{PT})$.

\begin{figure}[!h]
    \centering
    \includegraphics[width=400pt]{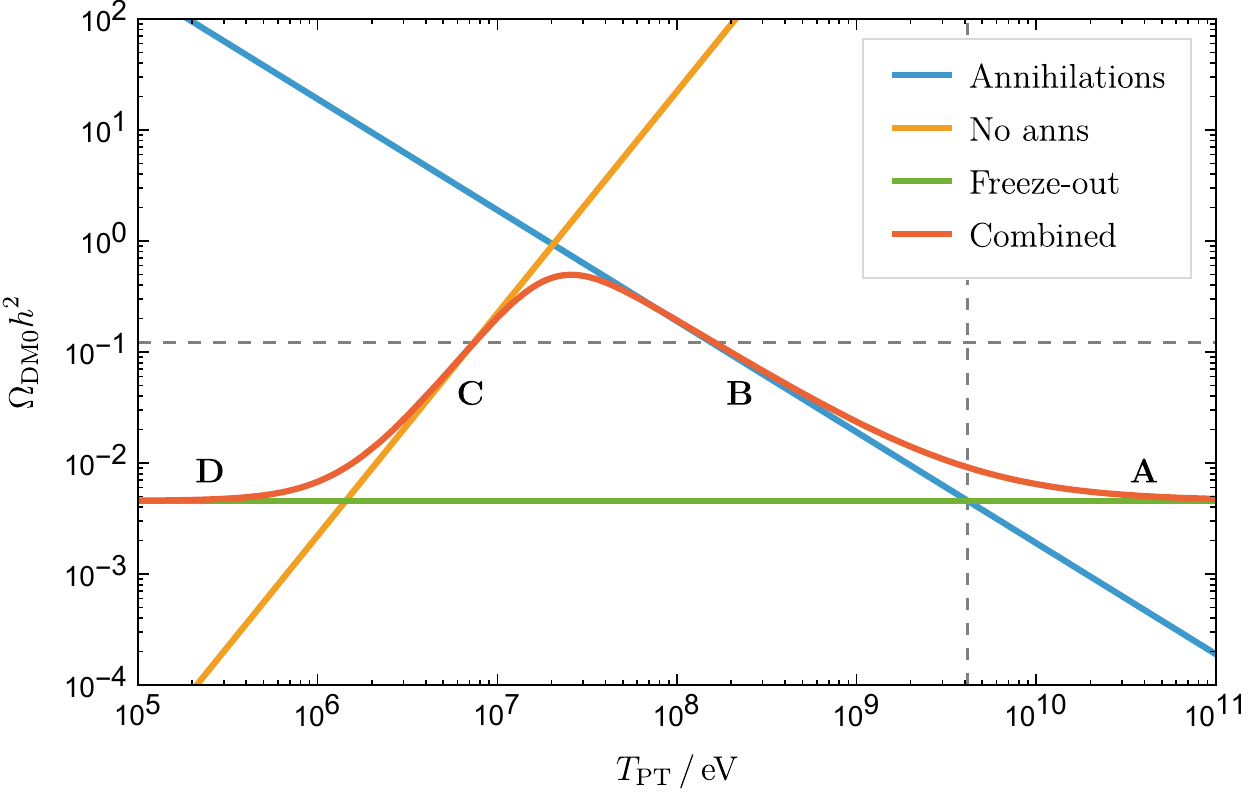}
    \caption{Final relic abundance of dark matter after the phase transition and potential subsequent annihilations, for values of $m_V = 100\,\rm GeV$, $\lambda = 0.1$, $\alpha, \xi = 1$. The green freeze out line is the relic abundance for these parameters assuming no phase transition. The horizontal dashed line is the observed dark matter abundance of $\Omega_{\rm DM0} h^2 = 0.121$, and the vertical represents the freeze-out temperature $T_\text{FO}$. The regions A, B, C \& D described in the text are labelled on the graph.}
    \label{fig:Omegah2}
\end{figure}

\begin{figure}
    \centering
    \includegraphics{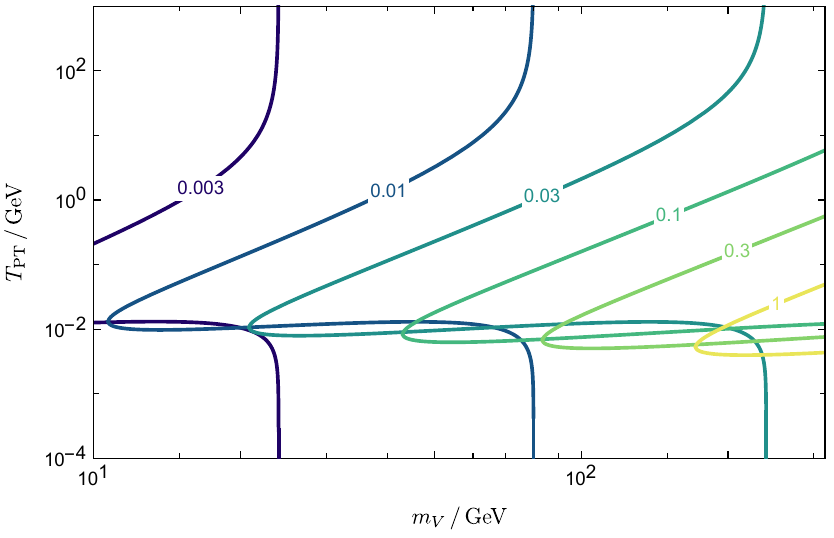}
    \caption{Vector-case solutions for $T_\text{PT}$ for varying values of $\lambda$ ($\alpha,\xi=1$).}
    \label{fig:TPTvec}
\end{figure}

\begin{figure}
    \centering
    \includegraphics{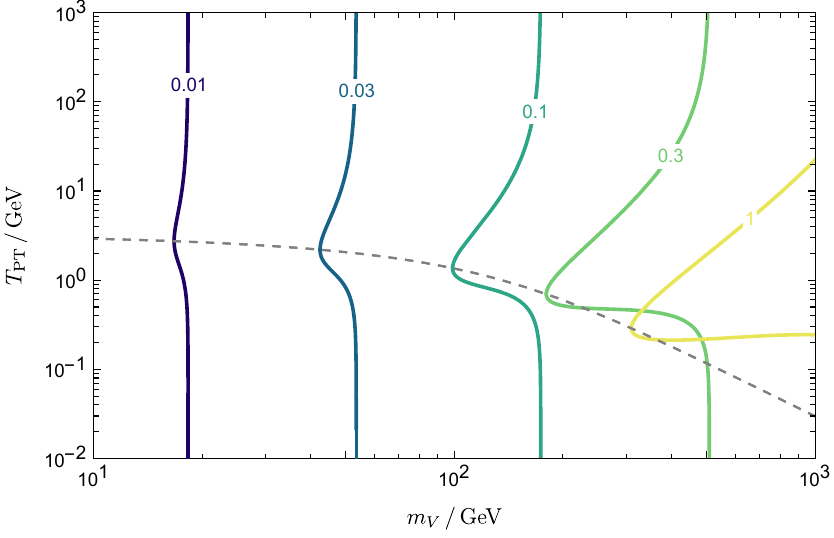}
    \caption{Scalar-case solutions for $T_\text{PT}$ for varying values of $\lambda$ ($\alpha,\xi=1$).}
    \label{fig:TPTscal}
\end{figure}

To produce the correct amount of dark matter in the annihilating case, we again solve Eq. \ref{relic_density} with $Y(\infty) = x_\text{PT}/\eta$, giving a contraint of
\begin{equation}
    T_\text{PT} = 0.16\times \bigg(\frac{100}{g_*}\bigg)^{\!\frac{1}{2}} \bigg(\frac{0.1}{\lambda}\bigg)^{\!2} \bigg(\frac{m_V}{100\,\text{GeV}}\bigg)^3\,\text{GeV}
\end{equation}
in the vector case, and in the scalar case,
\begin{equation}
    T_\text{PT} = 1.46\times \bigg(\frac{100}{g_*}\bigg)^{\!\frac{1}{2}} \bigg(\frac{0.1}{\lambda}\bigg)^{\!2} \bigg(\frac{m_V}{100\,\text{GeV}}\bigg)^3\,\text{GeV}
\end{equation}
Ordinary freeze-out of the vector dark matter will provide an additional contribution over that of the phase transition, given by $Y_\text{FO} = x_\text{FO}/\eta$, and will set a lower bound on the amount of dark matter produced. To account for this, we add it to the abundance given by the phase transition:
\begin{equation}
    Y(\infty) = Y_\text{FO} + \left(\frac{\eta}{x_\text{PT}} + \frac{1}{Y(x_\text{PT})}\right)^{-1}.
\end{equation}
There are now four regimes which characterise the evolution of the dark matter depending on the temperature at which the phase transition occurs.
\begin{enumerate}
  \renewcommand{\labelenumi}{\Alph{enumi}.}
  \item For phase transition temperatures larger than the freeze-out temperature, any dark matter produced by the phase transition quickly re-equilibrates before undergoing normal freeze-out, leaving no trace of the phase transition.
  \item For temperatures below freeze-out, the phase transition may still produce enough dark matter to reignite annihilations; most of the dark matter disappears, but a larger quantity than normal freeze-out remains.
  \item For yet lower phase transition temperatures, annihilations remain disfavoured and the raw quantity of dark matter produced by the phase transition (with a potential small contribution from freeze-out) is the amount that persists until today.
  \item At extremely low temperatures, the phase transition produces a negligible amount of dark matter and freeze-out dominates.
\end{enumerate}
These four regimes are demonstrated in Fig. \ref{fig:Omegah2}. Importantly, if the peak of this graph isn't high enough to reach the $\Omega_{\rm DM0} h^2 = 0.121$ line, then no solutions for the phase transition temperature exist that give the correct dark matter abundance. This limits the parameter space available to explore with this model for dark matter generation. Figs. \ref{fig:TPTvec} \& \ref{fig:TPTscal} show the range of phase transition temperature solutions for different values of the coupling $\lambda$ and dark matter masses.


\section{Gravitational Wave Production}
\label{sec:gw}

We wish to estimate the gravitational wave spectrum produced by the phase transition. As the bubble quickly reaches its terminal velocity, very little of the vacuum energy ends up in the bubble walls at collision, so their contribution to producing gravitational waves is very minimal. Initially, most of the vacuum energy resides in the momenta of the dark matter particles, but as they interact with the $\phi$-plasma, the momentum gets converted into bulk motion of the plasma i.e. sound waves. The timescale on which this occurs relative to the Hubble time is from Eq. \ref{trelax} found to be very fast. This would mean that the vacuum energy quickly ends up as sound waves in the plasma, putting us in a similar situation to that described in \cite{Caprini_2016, Hindmarsh_2015}, where a stochastic gravitational wave background is generated by bulk motion of the plasma after a first order phase transition has occurred. Following the fit to numerical results from \cite{Konstandin_2018}, the energy density of gravitational waves (with $v_w \approx 1$) is found to be
\begin{equation}
    \Omega_\text{gw}h^2 = 1.06\times10^{-6} \left(\frac{H_\text{PT}}{\beta}\right)^2 \left(\frac{\kappa \alpha}{1 + \alpha}\right)^2 \left(\frac{100}{g_* (T_\text{PT})}\right)^\frac{1}{3} S(f)\,,
\end{equation}
where $\kappa$ is the fraction of the vacuum energy transformed to sound waves in the plasma, which is taken to be 1, and $S(f)$ is the spectral shape function, normalised to 1 with a peak frequency of
\begin{equation}
    f_\text{peak} = 2.12\times10^{-5}\,\text{mHz}\,\left(\frac{\beta}{H_\text{PT}}\right) \left(\frac{T_\text{PT}}{\text{GeV}}\right) \left(\frac{g_* (T_\text{PT})}{100}\right)^\frac{1}{6}.
\end{equation}
The shape function takes the form
\begin{equation}
    S(f) = \frac{(a + b)f_\text{peak}^b f^a}{b f_\text{peak}^{a+b} + a f^{a+b}}\,,\quad\quad(a = 0.9, b = 2.1)\,.
\end{equation}
An additional IR cutoff imposed by causality, where the slope becomes $f^3$, is included for $f < H_\text{PT}/2\pi$.

\begin{figure}[!]
    \centering
    \includegraphics[width=415pt]{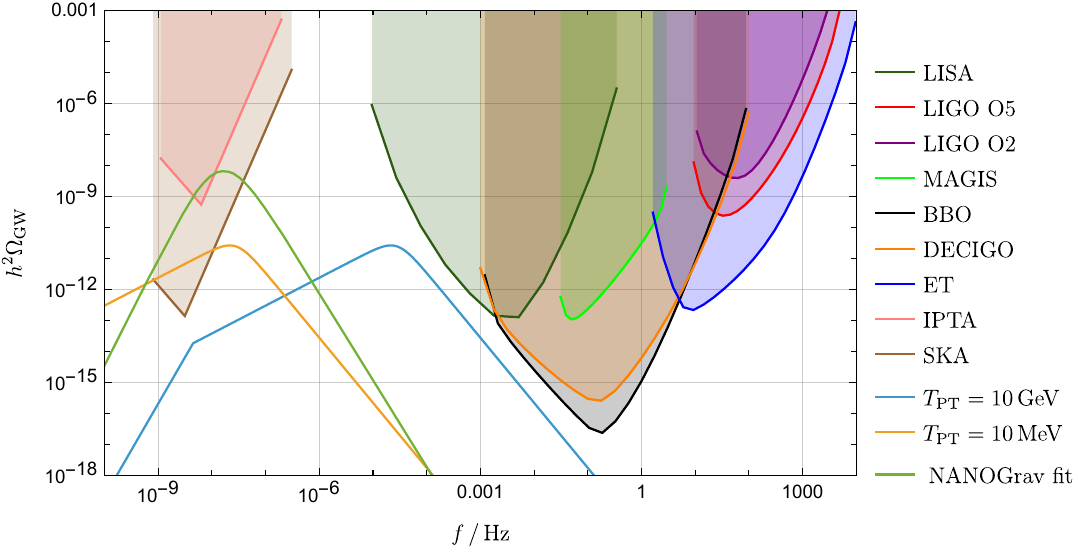}
    \caption{Gravitational wave signal (blue, orange) for phase transition parameters of $\alpha=1,\beta/H_\PT=100$ with sensitivity curves of current and planned gravitational wave observatories superimposed. The green line corresponds to the best fit to the NANOGrav detection if the signal is produced by a first-order phase transition, where a different model for the spectral function $S(f)$ from \cite{Caprini_2016} has been used.}
    \label{fig:gwsignal}
\end{figure}

\section{Conclusions}
In this work we have considered the production of both scalar and vector dark matter due to a first order phase transition where the particles are coupled to the field undergoing the phase transition.  We have revisited the equations pertaining to dark matter production in these situation and carefully followed the production of dark matter and its subsequent evolution, including annihilations, to explore the parameter space where a good relic abundance of dark matter can be obtained. 

We set up a scalar potential where there are minimal fine tunings between different scales like the latent heat, the vev in the broken phase and the critical temperature. We use dimensionless parameters $\alpha,\xi$ which can be tweaked should others wish to consider more fine tuned potentials.

We consider a given dark matter mass and coupling to the scalar field, choosing a combination of parameters that would not yield enough dark matter by thermal freeze-out alone.  We find there are then two different solutions for the phase transition temperature which yield the correct relic abundance, both of which lie below the freeze out temperature in the absence of phase transitions.  For the higher temperature phase transition, the particles are produced in a sufficiently high number density that annihilations restart again and subsequently play a role in the final relic abundance, although leaving an enhanced number density over the background freeze-out abundance.  For the lower temperature phase transition solution, the number density at that lower $T_\text{PT}$ isn't high enough for annihilation to be important and what is produced by the phase transition is precisely the number density required for good relic abundance.

Because the rate at which dark matter is produced depends upon the speed of the bubble wall, we also quantify the size of the region within each phase transition bubble where there might be different densities of dark matter, since such variations in density might conceivably lead to observable consequences.  Unfortunately we find that this is not the case and we predict no such observable effects, since the bubbles reach terminal velocity on a length scale many orders of magnitude smaller than the spatial separation of the bubble nucleation sites.

Some key conclusions are the following:-
\begin{itemize}
\item{For both the scalar and the vector case, there is a particular narrow range of phase transition temperatures where the relic abundance is altered significantly from the traditional freeze-out scenario, meaning different couplings and masses of dark matter to the background yield the correct amount of dark matter.}
\item{For vector dark matter, a phase transition around 10 MeV gives the biggest change from the normal freeze-out result.}
\item{For scalar dark matter, phase transitions between a few GeV and 100 MeV give rise to changes from normal freeze-out scenarios.}
\item{In both scenarios, the presence of a phase transition in the dark sector close to these range of masses changes the unitarity bound \cite{Griest:1989wd} and lowers the maximum mass of dark matter which can lead to relic abundance.}
\item{The same effect brings lower mass dark matter into a viable region of parameter space when the phase transition occurs at the correct temperature.}
\item{The gravitational wave signal of phase transitions that alter the relic abundance of dark matter would occur at low frequency and could be detected by future gravitational wave observations made by pulsar timing arrays.}
\end{itemize}

Strong first order phase transitions in a field coupled to dark matter can therefore change its abundance in quite significant and characteristic ways.  The phenomenology of such phase transitions depends on the coupling to the standard model, which in this paper we have assumed to be negligible so as to avoid significant friction on the bubble walls.  Characteristic gravitational wave signals are expected.

\acknowledgments
WS was supported by a UKRI STFC quota studentship.  MF was supported by the STFC under UKRI grant ST/X000753/1.  We are extremely grateful for conversations with Tevong You, Ken Mimasua and especially Wenyuan Ai.

\appendix

\section{Wall Friction and Dark Matter Production}
\label{sec:wallfric}

\subsection{Vector Dark Matter}

Taking $\vecp$ to be the momentum of the incoming $\phi$-particle in the wall frame, and $\veck_1,\,\veck_2$ to be the momenta of the produced $V$'s, we have $\Delta p^z = p^z - k_1^z - k_2^z$. In \cite{Ai:2024ikj}, the spin-averaged matrix element of such an interaction is calculated to be
\begin{equation}
    \sum_{r,r'}\lvert\mathcal{M}^{(r,\,r')}_{\phi\rightarrow VV}\rvert^2 = \frac{\lambda^2 v_b^2 L_w^2 \pi^2}{4}\left(2 + \frac{(k_1\cdot k_2)^2}{m_V^4}\right)\csch^2\left(\frac{\pi L_w \Delta p^z}{2}\right)\,,
\end{equation}
where $k_1,k_2$ are the 4-momenta of the produced $V$-particles. The middle term arises from the sum of polarisations, where the longitudinal mode allows for an enhancement in transition probability for large $k_1,k_2$. The $\csch^2$ term arises from the Fourier transform of the wall profile. The probability of transition is then
\begin{align}
\label{eq:dP}
    \mathbb{P}_{\phi\rightarrow  VV}(\vecp)=\frac{\lambda^2 v_b^2 L_{w}^2 \pi^2}{16 p^z}  &\prod_{i=1,2}\int\frac{\d^3\veck_i}{(2\pi)^3 2 E_{\veck_i}} (2\pi)^3 \delta(E_{\vecp}-E_{\veck_1}- E_{\veck_2})\delta^{(2)}(\vecp_\perp-\veck_{1,\perp}-\veck_{2,\perp})\notag\\
    &\times \left(2 + \frac{(k_1\cdot k_2)^2}{m_V^4}\right) {\rm csch}^2\left(\frac{\pi L_w \Delta p^z }{2}\right)\,,
\end{align}
Finally, to find the pressure on the wall, we integrate over the momentum $\vecp$ the product of the differential probability with the exchanged momentum $\Delta p^z$ and the flux of particles crossing the wall, which is the velocity ($\approx 1$) times the distribution function $f_\phi(\vecp, T)$ of $\phi$-particles in the frame of the wall.
\begin{equation}
    \mathcal{P}_{\phi\rightarrow  VV} = \int\frac{\d^3\vecp}{(2\pi)^3}\mathbb{P}_{\phi\rightarrow  VV}(\vecp)\times f_\phi(\vecp,T)\times \Delta p^z\,.
    \label{eq:pressure}
\end{equation}
As we assume an ultrarelativistic wall, with large $\gamma_w$, we may expand this complicated integral in $1/\gamma_w$ to simplify to an analytically tractable expression. First we consider the distribution of $\phi$'s in the boosted frame:
\begin{equation}
    f_\phi(\vecp,T) = \e^{-\frac{1}{T}\gamma_w\left(E^{(\phi)}_\vecp-v_w p^z\right)} \approx \e^{-\frac{1}{2T}\left(\frac{\gamma_w p_\perp^2}{p^z}+\frac{p^z}{\gamma_w}\right)}\,.
\end{equation}
$p_\perp$ is unboosted, so while $p^z\sim \gamma_w T$, $p_\perp\sim T$. In addition Ai (2023) \cite{Ai:2023suz} takes $L_w \sim O(10) / T$, so we see that $\Delta p^z \sim T$, otherwise the $\csch$ function becomes exponentially suppressed. This motivates expanding
\begin{equation}
    E_\veck \approx k^z + \frac{k_\perp^2 + m_V^2}{2 k^z}\,,
\end{equation}
as large $k_\perp$ implies large $\Delta p^z$. Taking $E_{\veck_1} = x E_\vecp$, we find that
\begin{equation}
    \Delta p^z \approx \frac{k_\perp^2 + m_V^2 + p_\perp^2 x(1 - x)}{2 p^z x(1 - x)}\,.
\end{equation}
$\Delta p^z$ does not strongly depend on $p_\perp\sim T\ll m_V$, so we neglect the $p_\perp$ term from now on. The $\csch$ term then becomes exponentially suppressed for $k_\perp \gg \sqrt{p^z / L_w}$. We similarly evaluate the polarisation term to be
\begin{equation}
    k_1\cdot k_2 \approx \frac{k_\perp^2 + m_V^2(1-2x+2x^2)}{2x(1-x)}\,.
\end{equation}
Taking $E_\vecp \approx p^z$ we arrive at the expression
\begin{align}
    \mathcal{P}_{\phi\rightarrow  VV} \approx \frac{\lambda^2 v_b^2 L_{w}^2 \pi^2}{16} &\int_0^\infty \frac{\d p^z}{(2\pi)^3} \int_0^1\frac{p^z \d x}{(2 \pi)^3} \frac{1}{4 (p^z)^3 x(1-x)} \e^{-\frac{p^z}{2T \gamma_w}} \notag\\\times&\int_0^\infty \d k_\perp 2 \pi k_\perp \frac{k_\perp^2}{2 p^z x(1 - x)} \frac{k_\perp^4}{4x^2(1-x)^2 m_V^4} \csch^2\left(\frac{\pi L_w}{2} \!\left( \frac{k_\perp^2}{2 p^z x(1 - x)}\right)\!\right) \notag\\\times&\int_0^\infty \d p_\perp 2\pi p_\perp \e^{-\frac{\gamma_w p_\perp^2}{2T p^z}}\,.
\end{align}
The terms involving $(k_\perp^2+m_V^2)$ have been simplified to $k_\perp^2$, as the integral is dominated by $k_\perp^2 \sim p^z / L_w \gg m_V^2$. This integral then evaluates to
\begin{equation}
    \mathcal{P}_{\phi\rightarrow  VV} \approx \frac{3\,\zeta(3)}{8 \pi^6} \frac{T_\text{PT}^4 v_b^2 \lambda^2}{L_w^2 m_V^4}\gamma_w^2\,,
\end{equation}
To calculate the number density $n_V$ of produced particles we simply drop the $\Delta p^z$ factor from the integrand, modifying the integral over $k_\perp$ to
\begin{equation}
    \int_0^\infty \d k_\perp 2 \pi k_\perp
    \frac{k_\perp^4}{4x^2(1-x)^2 m_V^4}
    \csch^2\left(\frac{\pi L_w}{2} \!\left( \frac{k_\perp^2}{2 p^z x(1 - x)}\right)\!\right)\,,
\end{equation}
which, when integrated as part of the full expression, and dividing by a factor of $\gamma_w$ to transform to the plasma frame gives the result
\begin{equation}
    n_V \approx \frac{T_\text{PT}^4 v_b^2 \lambda^2}{24 \pi^3 L_w m_V^4} \cdot \gamma_w\,.
\end{equation}
It will also be useful to know the energy of the produced particles in the plasma frame. For a rough estimate, we transform $E_\veck$ for $x=\tfrac{1}{2}$ back into the plasma frame, which gives approximately
\begin{equation}
    E_\veck^\text{plasma} \approx \gamma_w\frac{k_\perp^2}{4 p^z}\sim \frac{\gamma_w}{4 L_w}\,.
\end{equation}
More precisely, the energy density transferred to the plasma is equal to the friction $\mathcal{P}_{\phi\rightarrow  VV}$, with $\mathcal{P}_{\phi\rightarrow  VV}^\text{max}=\Delta V$ at the wall's terminal velocity i.e. all the wall's vacuum energy is going into producing dark matter, not speeding the wall up. Then the average energy per produced dark matter particle is
\begin{equation}
    \langle E_\veck^\text{plasma}\rangle = \frac{\mathcal{P}_{\phi\rightarrow  VV}}{n_V}\approx\frac{9\zeta(3)}{\pi^3 L_w}\gamma_w\,,
    \label{eq:dmenergy}
\end{equation}
which closely aligns with our previous result. As this is much larger than $m_V$, we have $p_V\approx E_\veck^\text{plasma}$, where we henceforth refer to $p_V$ as the dark matter momentum in the plasma frame. At the wall's terminal velocity, we have
\begin{equation}
    p_V^\text{max} = 6\sqrt{6 \zeta(3)}\frac{m_V^2\sqrt{\Delta V}}{T_\PT^2 v_b \lambda}\,.
\end{equation}

\subsection{Unitarity Violation}

As we have been working with a non-renormalisable Proca field, and haven't specified any UV completion, we must ensure that any unitarity limits aren't being violated, which we do by checking $\mathbb{P}_{\phi\rightarrow  VV}(\vecp) \leq 1$, which is calculated by evaluating the integral defined in Eq. \ref{eq:dP}, after making the simplifications specified above, 
\begin{align}
    \mathbb{P}_{\phi\rightarrow  VV}(\vecp) = \frac{\lambda^2 v_b^2 L_{w}^2 \pi^2}{16 p^z}&\int_0^1\frac{p^z \d x}{(2 \pi)^2} \frac{1}{4 (p^z)^3 x(1-x)} \notag\\\times&\int_0^\infty \d k_\perp 2 \pi k_\perp \frac{k_\perp^4}{4x^2(1-x)^2 m_V^4} \csch^2\left(\frac{\pi L_w}{2} \!\left( \frac{k_\perp^2}{2 p^z x(1 - x)}\right)\!\right)\,,
\end{align}
which gives the result
\begin{equation}
    \mathbb{P}_{\phi\rightarrow  VV}(\vecp) = \frac{T_\text{PT} v_b^2 \lambda}{96 \pi L_w m_V^4} \cdot \gamma_w\,.
\end{equation}
The transition probability increases linearly with $\gamma_w$, so will surpass a value of 1 at sufficiently large wall speeds. Substituting the expression for $\gamma_w^\mx$, and using the relations in Eq. \ref{eq:alpha} \& \ref{eq:xi} we find
\begin{equation}
    \mathbb{P}_{\phi\rightarrow  VV}^\mx(\vecp) = \frac{\pi^3 \sqrt{g_* \alpha} \xi \lambda}{144\sqrt{5\zeta(3)}} \frac{T_\PT^2}{m_V^2}\,.
\end{equation}
Thus the probability is suppressed by a factor of $(m_V/T_\PT)^2$ which, looking at the graph of solutions (Fig. \ref{fig:TPTvec}), is generally very large, and so we may be confident that unitarity is not violated in the regimes we are considering.

\subsection{Scalar Dark Matter}

To find the scalar friction, we drop the term arising from summing the polarisation states, giving the $k_\perp$ integral
\begin{equation}
    \int_0^\infty \d k_\perp 2 \pi k_\perp \frac{k_\perp^2 + m_\chi^2}{2 p^z x(1 - x)} \csch^2\left(\frac{\pi L_w}{2} \!\left( \frac{k_\perp^2 + m_\chi^2}{2 p^z x(1 - x)}\right)\!\right)\,.
\end{equation}
The integrand is largest for $k_\perp \sim m_\chi$ so we no longer may ignore the $m_\chi^2$ terms. This integrates to
\begin{equation}
    \mathcal{P}_{\phi\rightarrow \chi\chi} = \frac{T_\text{PT}^2 v_b^2 \lambda^2}{128\pi^4}\left(-1-\gamma_E+\log\left(\frac{4T_\text{PT}\gamma_w}{\pi L_w m_\chi^2}\right)\right)\,.
\end{equation}
Finally, the $k_\perp$ integral for the scalar number density is
\begin{equation}
    \int_0^\infty \d k_\perp 2 \pi k_\perp \csch^2\left(\frac{\pi L_w}{2} \!\left( \frac{k_\perp^2 + m_\chi^2}{2 p^z x(1 - x)}\right)\!\right)\,,
\end{equation}
giving
\begin{equation}
    n_\chi = \frac{T_\text{PT}^3 v_b^2 \lambda^2}{48\pi^4 m_\chi^2}\,.
\end{equation}
The final momentum of the scalar particles is, using Eq. \ref{eq:dmenergy}
\begin{equation}
    p_\chi = \frac{3m_\chi^2}{8T_\PT}\left(-1-\gamma_E+\log\left(\frac{4T_\text{PT}\gamma_w}{\pi L_w m_\chi^2}\right)\right)\,.
\end{equation}

\section{Observational Effects}
\label{sec:obs}

We are interested in seeing if this method of generating dark matter via a phase transition leads to any deviation from the behaviour of dark matter in the traditional freeze-out scenario, in particular, whether it could lead to inhomogeneities in the dark matter distribution that collapse out of the expansion of the Universe, leaving behind compact subhaloes that could be detected via weak lensing and stellar stream methods \cite{Banik_2021}. One effect to consider is that the bubble takes time to speed up to its terminal velocity, and during this phase the amount of dark matter produced increases with distance from the site of bubble nucleation. The radius of the bubble at percolation is
\begin{equation}
    r_\text{perc} = \frac{(8\pi)^\frac{1}{3}}{(\beta/H_\text{PT})\,H_\text{PT}}\,,
\end{equation}
while the radius at which the bubble reaches its terminal velocity is
\begin{equation}
    r_{\rm term} = \frac{v_b^2}{3 L_w \sqrt{\mathcal{C} \Delta V}} = 1.72\times10^{-27}\,\text{pc} \times \frac{\xi}{\sqrt{\alpha}}\bigg(\frac{100}{g_*}\bigg)^{\!\frac{1}{2}} \bigg(\frac{0.1}{\lambda}\bigg) \bigg(\frac{m_V}{100\,\text{GeV}}\bigg)^2 \bigg(\frac{\rm GeV}{T_{\rm PT}}\bigg)^3.
\end{equation}
This indicates that generally the bubble reaches its terminal velocity extremely quickly, and then produces a uniform dark matter distribution from then on.
\begin{figure}[!]
    \centering
    \includegraphics[width=400pt]{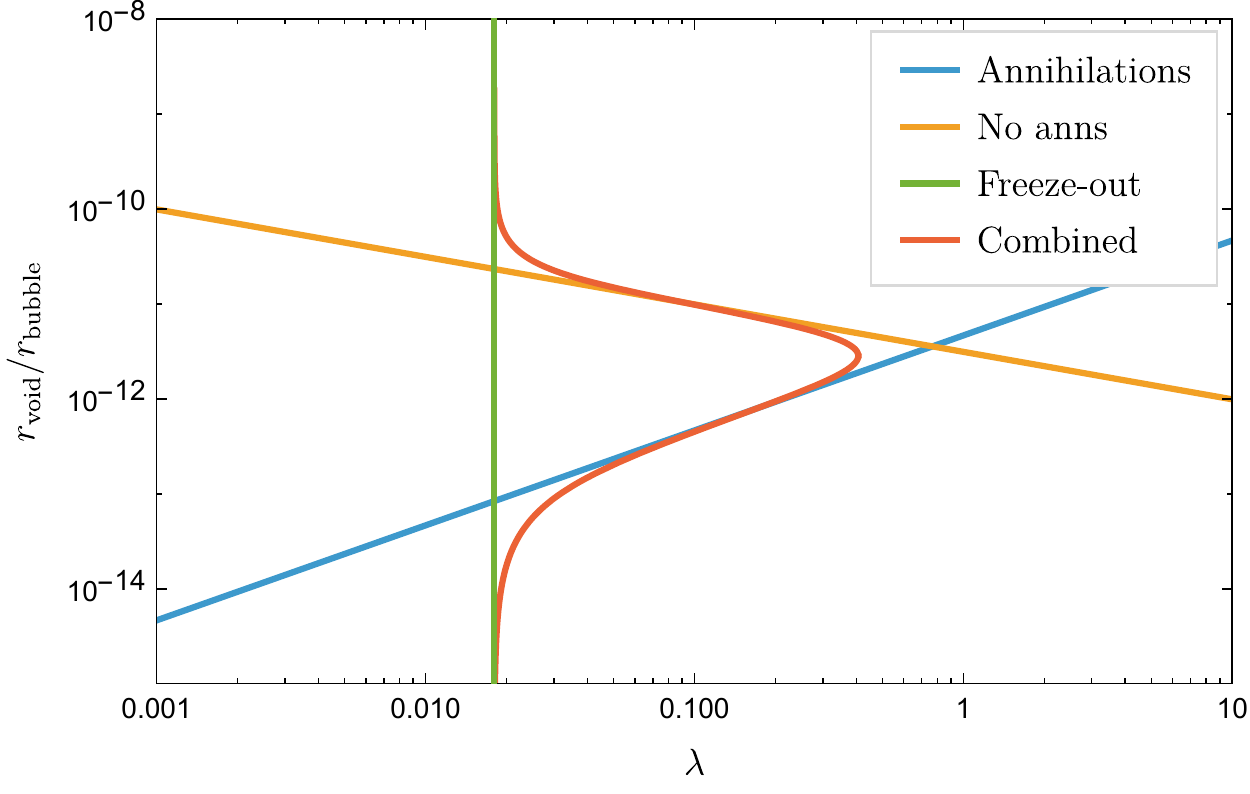}
    \caption{Ratio of void to bubble for $m_V = 1\,\rm TeV$, $\kappa, \xi = 1$.}
    \label{fig:ratio}
\end{figure}
Another effect that will be considered is the behaviour of the bubbles when they collide, as this has been proposed as a mechanism for generating dark matter in its own right. This could lead to overdensities in the dark matter distribution near the boundaries of the bubbles. However, as we know that the bubble reaches its terminal velocity quickly, the wall does not carry much energy in comparison to the total energy released by the false vacuum, so contributions to the dark matter distribution from wall collisions are expected to be minimal.


\bibliographystyle{JHEP}
\bibliography{biblio.bib}






\end{document}